\documentclass[10pt,letterpaper]{article}
\usepackage{opex3}
\usepackage{color}
\usepackage{cite}
\usepackage{amsmath,xspace}
\usepackage{amssymb}

\usepackage[normalem]{ulem}

\begin{document}
% additional commands PMP
\newcommand{\Rb}{$^{87}$Rb\xspace}
\newcommand{\angf}{2 \pi \times}
\newcommand{\insitu}{\textit{in situ}}
\newcommand{\unsim}{\mathord{\sim}} %smaller space after sim

%%%%%%%%%%%%%%%%%% title page information %%%%%%%%%%%%%%%%%%
%\title{Holographic beam shaping for single-site addressing of ultracold atoms in optical lattices}
\title{Ultra-precise holographic beam shaping for microscopic quantum control}
\author{Philip Zupancic$^{1,2}$, Philipp M. Preiss$^{1,3}$, Ruichao Ma$^{1,4}$, Alexander Lukin$^1$, M. Eric Tai$^1$, Matthew Rispoli$^1$, Rajibul Islam$^1$, and Markus Greiner$^{1*}$}

\address{$^1$Department of Physics, Harvard University, Cambridge,
Massachusetts, 02138, USA\\
$^2$Present address: Institute for Quantum Electronics, ETH Z\"{u}rich,
8093 Z\"{u}rich, Switzerland\\
$^3$Present address: Physikalisches Institut, Heidelberg University,
69120 Heidelberg, Germany\\
$^4$Present address: James Franck Institute, Department of Physics,
The University of Chicago, Chicago, Illinois 60637}

\email{$^*$greiner@physics.harvard.edu} %% email address is required

%%%%%%%%%%%%%%%%%%% abstract and OCIS codes %%%%%%%%%%%%%%%%
%% [use \begin{abstract*}...\end{abstract*} if exempt from copyright]

\begin{abstract}
High-resolution addressing of individual ultracold atoms, trapped ions or solid state emitters allows for exquisite control in quantum optics experiments. This becomes possible through  large aperture magnifying optics that project microscopic light patterns with diffraction limited performance. We use programmable amplitude holograms generated on a digital micromirror device to create arbitrary microscopic beam shapes with full phase and amplitude control. The system self-corrects for aberrations of up to several $\lambda$ and reduces them to $\lambda/50$, leading to light patterns with a precision on the $10^{-4}$ level. We demonstrate aberration-compensated beam shaping in an optical lattice experiment and perform single-site addressing in a quantum gas microscope for $^{87}$Rb.
\end{abstract}

\ocis{(070.6120) Spatial light modulators, (090.1000) Holography: Aberration compensation, (090.1995) Digital holography, (140.7010) Laser trapping, (350.4855) Optical tweezers or optical manipulation, (020.1475) Bose-Einstein condensates} % REPLACE WITH CORRECT OCIS CODES FOR YOUR ARTICLE, MINIMUM OF TWO; Avoid using the OCIS codes for “General” or “General science” whenever possible.

%%%%%%%%%%%%%%%%%%%%%%% References %%%%%%%%%%%%%%%%%%%%%%%%%

\bibliographystyle{osajnl}
\bibliography{bibs}

\begin{thebibliography}{10}
\newcommand{\enquote}[1]{``#1''}

\bibitem{Jurcevic2014}
P.~Jurcevic, B.~P. Lanyon, P.~Hauke, C.~Hempel, P.~Zoller, R.~Blatt, and C.~F.
  Roos, \enquote{Quasiparticle engineering and entanglement propagation in a
  quantum many-body system,} Nature \textbf{511}, 202--205 (2014).

\bibitem{Weitenberg2011}
C.~Weitenberg, M.~Endres, J.~F. Sherson, M.~Cheneau, P.~Schau\ss, T.~Fukuhara,
  I.~Bloch, and S.~Kuhr, \enquote{Single-spin addressing in an atomic {Mott}
  insulator,} Nature \textbf{471}, 319--324 (2011).

\bibitem{Wang2015}
Y.~Wang, X.~Zhang, T.~A. Corcovilos, A.~Kumar, and D.~S. Weiss,
  \enquote{Coherent addressing of individual neutral atoms in a {3D} optical
  lattice,} Phys. Rev. Lett. \textbf{115}, 043003 (2015).

\bibitem{Kurtsiefer2000}
C.~Kurtsiefer, S.~Mayer, P.~Zarda, and H.~Weinfurter, \enquote{Stable
  solid-state source of single photons,} Phys. Rev. Lett. \textbf{85}, 290--293
  (2000).

\bibitem{Boyer2006}
V.~Boyer, R.~M. Godun, G.~Smirne, D.~Cassettari, C.~M. Chandrashekar, A.~B.
  Deb, Z.~J. Laczik, and C.~J. Foot, \enquote{Dynamic manipulation of
  {Bose-Einstein} condensates with a spatial light modulator,} Phys. Rev. A
  \textbf{73}, 031402 (2006).

\bibitem{Gaunt2012}
A.~L. Gaunt and Z.~Hadzibabic, \enquote{Robust digital holography for ultracold
  atom trapping,} Sci. Rep. \textbf{2}, 721 (2012).

\bibitem{Nogrette2014}
F.~Nogrette, H.~Labuhn, S.~Ravets, D.~Barredo, L.~B\'eguin, A.~Vernier,
  T.~Lahaye, and A.~Browaeys, \enquote{Single-atom trapping in holographic {2D}
  arrays of microtraps with arbitrary geometries,} Phys. Rev. X \textbf{4},
  021034 (2014).

\bibitem{Beckers1993}
J.~M. {Beckers}, \enquote{{Adaptive optics for astronomy - Principles,
  performance, and applications},} Annual Review of Astronomy and Astrophysics
  \textbf{31}, 13--62 (1993).

\bibitem{Reicherter:99}
M.~Reicherter, T.~Haist, E.~U. Wagemann, and H.~J. Tiziani, \enquote{Optical
  particle trapping with computer-generated holograms written on a
  liquid-crystal display,} Opt. Lett. \textbf{24}, 608--610 (1999).

\bibitem{Wulff:06}
K.~D. Wulff, D.~G. Cole, R.~L. Clark, R.~DiLeonardo, J.~Leach, J.~Cooper,
  G.~Gibson, and M.~J. Padgett, \enquote{Aberration correction in holographic
  optical tweezers,} Opt. Express \textbf{14}, 4170--4175 (2006).

\bibitem{booth2002adaptive}
M.~J. Booth, M.~A. Neil, R.~Ju{\v{s}}kaitis, and T.~Wilson, \enquote{Adaptive
  aberration correction in a confocal microscope,} Proceedings of the National
  Academy of Sciences \textbf{99}, 5788--5792 (2002).

\bibitem{Vellekoop2008}
I.~M. {Vellekoop} and A.~P. {Mosk}, \enquote{{Universal Optimal Transmission of
  Light Through Disordered Materials},} Physical Review Letters \textbf{101},
  120601 (2008).

\bibitem{mosk2012controlling}
A.~P. Mosk, A.~Lagendijk, G.~Lerosey, and M.~Fink, \enquote{Controlling waves
  in space and time for imaging and focusing in complex media,} Nature
  photonics \textbf{6}, 283--292 (2012).

\bibitem{Lee:74}
W.-H. Lee, \enquote{Binary synthetic holograms,} Appl. Opt. \textbf{13},
  1677--1682 (1974).

\bibitem{Goorden:14}
S.~A. Goorden, J.~Bertolotti, and A.~P. Mosk, \enquote{Superpixel-based spatial
  amplitude and phase modulation using a digital micromirror device,} Opt.
  Express \textbf{22}, 17999--18009 (2014).

\bibitem{Schine2015}
N.~Schine, A.~Ryou, A.~Gromov, A.~Sommer, and J.~Simon, \enquote{Synthetic
  {L}andau levels for photons,} arXiv 1511.07381  (2015).

\bibitem{Papageorge2016}
A.~T. Papageorge, A.~J. Koll\'{a}r, and B.~L. Lev, \enquote{Coupling to modes
  of a near-confocal optical resonator using a digital light modulator,} Opt.
  Express \textbf{24}, 11447--11457 (2016).

\bibitem{Bakr2009}
W.~S. Bakr, J.~I. Gillen, A.~Peng, S.~F\"{o}lling, and M.~Greiner, \enquote{A
  quantum gas microscope for detecting single atoms in a {H}ubbard-regime
  optical lattice,} Nature \textbf{462}, 74--77 (2009).

\bibitem{Sherson2010}
J.~F. Sherson, C.~Weitenberg, M.~Endres, M.~Cheneau, I.~Bloch, and S.~Kuhr,
  \enquote{Single-atom resolved fluorescence imaging of an atomic {M}ott
  insulator,} Nature \textbf{467}, 68--72 (2010).

\bibitem{Preiss2015}
P.~M. Preiss, R.~Ma, M.~E. Tai, A.~Lukin, M.~Rispoli, P.~Zupancic, Y.~Lahini,
  R.~Islam, and M.~Greiner, \enquote{Strongly correlated quantum walks in
  optical lattices,} Science \textbf{347}, 1229--1233 (2015).

\bibitem{Islam2015}
R.~Islam, R.~Ma, P.~M. Preiss, M.~E. Tai, A.~Lukin, M.~Rispoli, and M.~Greiner,
  \enquote{Measuring entanglement entropy in a quantum many-body system,}
  Nature \textbf{528}, 77--83 (2015).

\bibitem{Zupancic2013}
P.~Zupancic, \enquote{Dynamic holography and beamshaping using digital
  micromirror devices,} Master's thesis, Ludwig-Maximilians-Universit\"at
  M\"unchen (2013).

\bibitem{Cizmar2010}
\v{C}i\v{z}m\'{a}r T., M.~Mazilu, and K.~Dholakia, \enquote{\textit{In situ}
  wavefront correction and its application to micromanipulation,} Nature
  Photon. \textbf{4}, 388--394 (2010).

\bibitem{Deutsch2008}
B.~Deutsch, R.~Hillenbrand, and L.~Novotny, \enquote{Near-field amplitude and
  phase recovery using phase-shifting interferometry,} Opt. Express
  \textbf{16}, 494--501 (2008).

\bibitem{Preiss2015thesis}
P.~M. Preiss, \enquote{Atomic {B}ose-{H}ubbard systems with single-particle
  control,} Ph.D. thesis, Harvard University (2015).

\bibitem{Bakr2010}
W.~S. Bakr, A.~Peng, M.~E. Tai, R.~Ma, J.~Simon, J.~I. Gillen, S.~F\"{o}lling,
  L.~Pollet, and M.~Greiner, \enquote{Probing the superfluid--to--{M}ott
  insulator transition at the single-atom level,} Science \textbf{329},
  547--550 (2010).

\bibitem{Kantian2014}
A.~Kantian, U.~Schollw\"ock, and T.~Giamarchi, \enquote{Competing regimes of
  motion of {1D} mobile impurities,} Phys. Rev. Lett. \textbf{113}, 070601
  (2014).

\bibitem{Smacchia2015}
P.~Smacchia, M.~Knap, E.~Demler, and A.~Silva, \enquote{Exploring dynamical
  phase transitions and prethermalization with quantum noise of excitations,}
  Phys. Rev. B \textbf{91}, 205136 (2015).

\bibitem{Fukuhara2015}
T.~Fukuhara, S.~Hild, J.~Zeiher, P.~Schau\ss{}, I.~Bloch, M.~Endres, and
  C.~Gross, \enquote{Spatially resolved detection of a spin-entanglement wave
  in a {B}ose-{H}ubbard chain,} Phys. Rev. Lett. \textbf{115}, 035302 (2015).

\bibitem{Kleine2008}
A.~Kleine, C.~Kollath, I.~P. McCulloch, T.~Giamarchi, and U.~Schollw\"ock,
  \enquote{Spin-charge separation in two-component {Bose} gases,} Phys. Rev. A
  \textbf{77}, 013607 (2008).

\end{thebibliography}
%%%%%%%%%%%%%%%%%%%%%%%%%%  body  %%%%%%%%%%%%%%%%%%%%%%%%%%

\section{Introduction} %PMP and PPJZ

Applications in quantum optics frequently rely on the interaction of light with quantum particles. In the highly controlled and isolated systems that are now available on a variety of platforms, it is often possible to individually address quantum objects, for example cold trapped ions \cite{Jurcevic2014}, ultracold atoms in optical lattices \cite{Weitenberg2011, Wang2015}, or emitters in solid-state environments \cite{Kurtsiefer2000}. To this end, it is necessary to create high-precision optical beam shapes, with complete control over beam profiles, alignment and focus \cite{Boyer2006,Gaunt2012,Nogrette2014}. 

In many experimental settings, achieving optimal beam shaping is hindered by aberrations in high-resolution imaging systems and the difficulty to reliably align and focus optical beams on the target, which may be embedded in a solid state environment or trapped inside a vacuum chamber. Methods for aberration compensation and adaptive optics have rapidly developed in the last decades, most prominently in the fields of astronomy \cite{Beckers1993}, optical tweezers \cite{Reicherter:99, Wulff:06} and microscopy \cite{booth2002adaptive, Vellekoop2008, mosk2012controlling}. %This task is even more challenging if beam profiles other than Gaussian are needed.

Here we present a technique for holographic beam shaping that is optimally suited for high-precision optical control in almost any quantum optics setting. By using a digital micromirror device (DMD) as a programmable amplitude hologram in a Fourier plane, we are able to engineer wavefronts with both phase and amplitude control \cite{Lee:74, Goorden:14,Schine2015, Papageorge2016}. We demonstrate the generation of high-order Laguerre- and Hermite-Gauss beams, as well as arbitrary, numerically defined beam profiles.

Using the complete control over the wavefront, it is possible to correct for known aberrations and achieve diffraction-limited performance even with phase errors of several $\lambda$ in the imaging system. The beam-shaping system itself can be used to measure aberrations in the imaging system, and to pre-correct the engineered wavefront accordingly. Here, we use a photodiode as an intensity point detector in an image plane to map out aberrations, which automatically guarantees diffraction-limited performance, as well as alignment and focus onto the point detector. We obtain exceptionally precise optical potentials with four orders of magnitude of dynamic range and wavefront errors as low as $\lambda/50$. Similar performance can be achieved for any quantum object that can act as a point intensity probe, for example trapped ions or solid state emitters via their fluorescence rate. As an application, we demonstrate how to achieve diffraction-limited arbitrary beam profiles in an optical lattice experiment with single-site resolution \cite{Bakr2009, Sherson2010}. We use DMD-generated beams to address individual atoms in an optical lattice and manipulate their dynamics. 

Holographically defined potentials afford novel capabilities in quantum gas microscopes and have allowed us to observe two-particle quantum walks \cite{Preiss2015} and to detect entanglement entropy in Bose-Hubbard systems \cite{Islam2015}. Similar approaches to DMD-based holography have recently been developed for mode-matched light injection into optical resonators \cite{Schine2015, Papageorge2016}.

\section{Holography with digital micromirror devices} \label{sec:holography}%PPJZ

We use the digital micromirror device DLP5500 from Texas Instruments as a spatial light modulator. This optomechanical device consists of $1024\times768$ square mirrors of width $10.8 \,\mu$m that are individually mounted on diagonal torsion hinges on top of an array of CMOS memory cells. Each mirror can be switched between a $+12^\circ$ ``on" and a $-12^\circ$ ``off'' orientation, realizing a binary intensity filter with switching frequencies up to 5\,kHz. Other versions with more than 4 megapixels and refresh rates of 50\,kHz are available. Unlike for LCD-based spatial light modulators, which require periodic reversals of the pixel-defining field, binary patterns displayed on the DMD are completely static and avoid fluctuations due to intensity modulation.

\begin{figure}
\begin{center}
\includegraphics[width=0.9\linewidth]{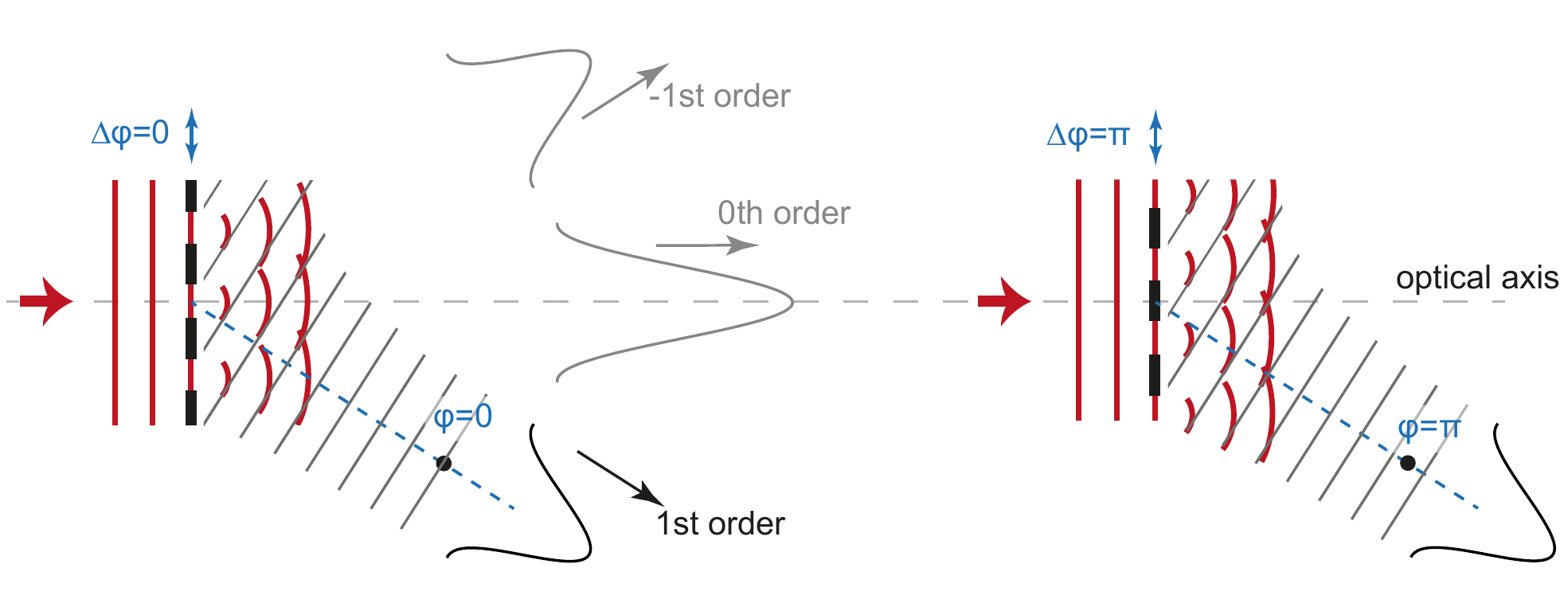}
\caption{Phase-shifting with an amplitude hologram. The phase of the light in the first diffraction order is given by the spatial phase of the grating. Thus we can change the phase of the light by modifying the position of the grating slits. This is illustrated here by the locations of the wave fronts relative to a fixed reference point (black dot) in the path of the beam.}
\label{fig:hologram}
\end{center}
\end{figure}

We employ the DMD holographically in the Fourier plane, which enables both local amplitude and phase control despite the direct DMD modulations being binary in intensity. In order to illustrate the idea, consider a diffraction grating of fixed periodicity (set to 1 in the following). The $m^\textrm{th}$ diffraction order lies in the direction where the phase difference between neighboring slits is $2\pi m$, or the light emerging from the grating has a linear phase gradient of slope $2\pi m / \mathrm{slit}$ (Fig.\ \ref{fig:hologram}). The outgoing light field of a single slit of width $a$ (a fraction of 1) centered around $x_0$ is given by
	\begin{align}
	\label{eqampintegral}
	E_\mathrm{out} &= E_\mathrm{in} \int\limits_{x_0-a/2}^{x_0+a/2} \exp (i\,2\pi m x) \, \mathrm{d} x \nonumber\\
	&= E_\mathrm{in} \frac{\sin(\pi m a)}{\pi m} \exp(i\, 2\pi m x_0).
	\end{align}
The phase $\phi$ of the outgoing field is linearly dependent on the position of the slit, $\phi\sim mx_0$. We use this to imprint a phase modulation onto the beam by choosing the correct slit position. The intensity of the light field depends on the slit width $a$, and the diffraction orders have relative amplitudes $\sin(\pi m a)/\pi m$. The  $0^\mathrm{th}$ diffraction order of the displayed grating cannot be used for phase control since it imparts a constant phase over the entire area, and we use the $+1^\mathrm{st}$ order for beam shaping.

We obtain the desired phase and amplitude profiles by displaying a grating with a period $p$ of e.g.\ 4 pixels at an angle of $\theta$ on the DMD and locally changing the slit widths and positions (examples in Fig.\ \ref{fig:beamshaping1}). The DMD is illuminated with monochromatic light of wavelength $\lambda$. Assuming we have an incoming amplitude profile $E_\mathrm{in}(x,y)$ with phase profile $\phi_\mathrm{in}(x,y)$, and we want to have an amplitude profile $E_\mathrm{out}(x,y)$ and a phase profile  $\phi_\mathrm{out}(x,y)$, the DMD pattern is a binarized version of
	 \begin{equation} 
	 \label{eqgrating}
	 \frac12 \left[1+\cos\left(\frac{2\pi}p [\cos\theta \cdot x + \sin\theta \cdot y] +\vphantom{\frac12}  \phi_\mathrm{out}(x,y) -  \phi_\mathrm{in}(x,y)\right)\right] 
	 \end{equation}
with a local ``on" fraction given by the slit width
	\begin{equation}
	a(x, y) = \frac1\pi \mathrm{arcsin} \left( s \pi \frac {E_\mathrm{out}(x,y)} {E_\mathrm{in}(x,y)} \right)
	\label{eqslitwidth}
	\end{equation}
from eq.\ \ref{eqampintegral}, where the scaling factor $s$ is given by  $s=(\min{(E_\mathrm{in}(x,y)/E_\mathrm{out}(x,y))})^{-1}$. $E_\mathrm{out}$ is the Fourier transform of the desired shape in the image plane.

We compared two methods for the generation of binary holograms from the smooth phase and amplitude profiles: A completely deterministic algorithm, i.e.\ rounding eq.\ \ref{eqgrating} to either 0 or 1, is prone to artifacts which occur if the effective periodicity of the grating is locally close to an integer such that it matches the periodicity of the mirror array. Every pixel has a grating phase assigned, and this phase determines if the pixel is on or off. If the periods of the DMD mirrors and the holographic pattern match, some phases occur more often than others, thus sampling the grating in a biased way --- a bias towards 0 over-represents ``on''-pixels while one towards $\pi/2$ leads to too many dark pixels. DMD areas with artifacts reduce the ability to define the local amplitude. This effect is enhanced if $\theta$, the angle of the pattern on the DMD, is chosen too small, such that the grating is aligned with the DMD. A randomized binarization that assigns ``on'' to a pixel with a higher probability the greater the value of eq.\ \ref{eqgrating}, can reduce artifacts. Choosing $\theta=10^\circ$ and employing a randomized algorithm yields good results \cite{Zupancic2013}.

While the holographic DMD setup allows for phase and amplitude control, a down-side of this configuration is its low efficiency in terms of laser power. The first non-trivial diffraction order $m=\pm1$ carries only $1/\pi^2\approx10\%$ of the incoming intensity. This grating sits on top of an inherent secondary grating formed by the individual mirrors, and the reflected light is distributed among several dozens of diffraction orders. Power can be maximized in one diffraction order by fulfilling the blazing condition, i.e. finding a simultaneous solution to both the grating equation and the law of reflection. We were able to get 55\% into one diffraction order of the mirror grating, which leaves us with 5.6\% in the first order of the displayed pattern. Modulations of the amplitude profile reduce the power of the resulting beam even further to typically 1--2\%.

\section{In situ calibration procedure} \label{sec:calibration}%PPJZ

Optical high-precision applications demand the absence of aberrations in the optical setup. Aberrations manifest themselves as deformations in the phase front of a beam or an image. Similar to the approach in references \cite{Papageorge2016, Cizmar2010}, the holographic employment of the DMD allows manipulations on the local phase and can be utilized to correct image defects if their parameters are known.
	
\begin{figure}
\begin{center}
\includegraphics[width=0.9\linewidth]{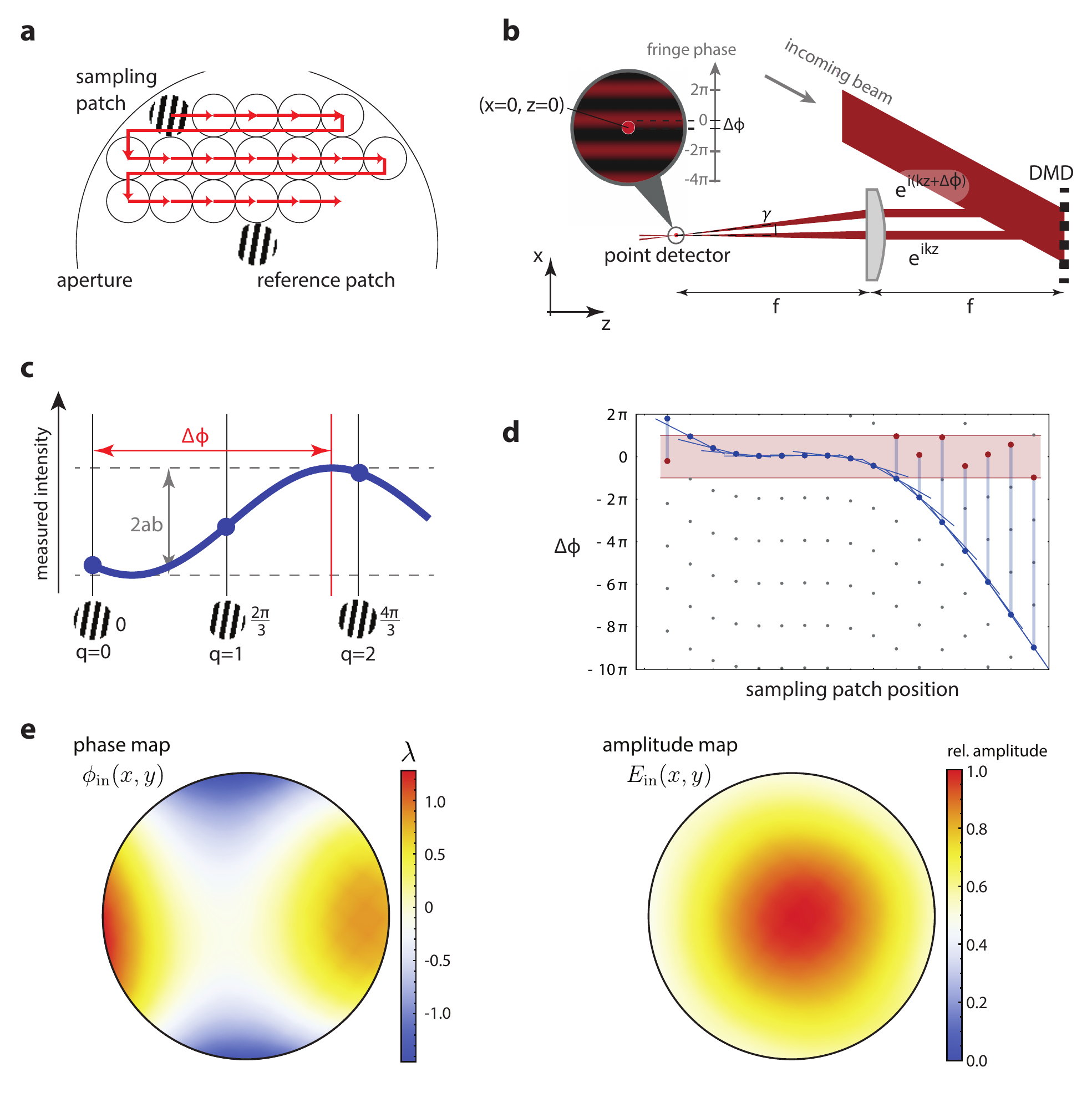}
\caption{Self-calibration of the aberration correction. (a) In the calibration sequence, only two small regions of the DMD are active --- one is fixed, the other samples the full aperture. (b) The DMD is placed in the Fourier plane of the optical system. Light emerging from the two patches interferes in the image plane. The position of the interference stripes depends on the relative phase between the two beams. (c) Using the phase-shifting capabilities of the holographic DMD, we sample the interference patterns at the same point in space for three different relative phases. From this data, the phase difference between the beams and their relative amplitude can be derived. (d) The measured phase difference is $\mathrm{mod}\ 2\pi$ (shaded interval). Adding integer multiples of $2\pi$ (gray dots), the true, smoothly varying phase profile can be found by unwrapping (blue dots). (e) Phase and amplitude map of the fully calibrated DMD path. Tilt and focus have been removed from the phase map, showing 3$\lambda$ of astigmatism due to bending of the DMD chip itself. The amplitude map shows the Gaussian envelope of the illuminating beam, which falls to 50\% of its maximal value at the edge of the aperture.}
\label{fig:calibration}
\end{center}
\end{figure}  	
	
Aberrations in an imaging system can be characterized using the DMD itself by displaying a sequence of patterns and recording the response in the image plane. Our scheme requires the direct or indirect measurement of the light intensity in the image plane.

A pattern sequence shows diffraction gratings on two small regions of 40 pixels diameter. One is located in the center (\textit{reference patch}), the other one is moved between measurements to cover the entire area of the DMD (\textit{sampling patch}, Fig.\ \ref{fig:calibration}(a)). The grating phase of the sampling patch can be increased in steps of $2\pi/3$, labeled by the index $q$. The two beams emerging from those regions will overlap in the image plane and form a standing wave interference pattern, whose spatial phase is equal to the phase difference between the two beams (Fig.\ \ref{fig:calibration}(b)). The beams converging towards the focal plane at an angle $\gamma$ can be described by
    \begin{eqnarray*}
    B_\mathrm{ref}(x, z) &=& a \cdot \exp\left(i k z\right) \\
    B_\mathrm{samp,q}(x, z) &=& b \cdot \exp\left(ik [\cos \gamma \cdot z-\sin \gamma \cdot x + \Delta\phi+q (2 \pi/3)]\right),
    \end{eqnarray*}
    where $k=2\pi/\lambda$, such that the intensity of the interference pattern becomes
    \begin{equation*}
    |B_\mathrm{ref}(x, z)+B_\mathrm{samp,q}(x, z)|^2 = a^2 + b^2 + 2ab \cdot \cos \left(k[ (1-\cos \gamma) z + \sin \gamma \cdot x ] + \Delta\phi +q (2 \pi/3)\right).
    \end{equation*}

Thus, the interference fringes at the origin  $(x=0, z=0)$ have a contrast of $2ab$ and a phase of $\Delta\phi$, which can be measured using a point detector, such as a photodiode behind a pinhole, a small region on an image sensor, or a single fluorescent atom, ion or solid state emitter. By making three measurements with $q=0,1,2$, adding $2\pi/3$ to $\Delta\phi$ each time by changing the phase of the reference patch and hence translating the interference pattern, $ab$ and $\Delta\phi$ can be determined (Fig.\ \ref{fig:calibration}(c)). From the three intensity measurements $m_q$, the phasor\cite{Deutsch2008}
\begin{equation*}
p = - \frac13 (m_2+m_3-2m_1) +  \frac i{\sqrt3} (m_2-m_3),
\end{equation*}
can be used to calculate their values:
\begin{align*}
\Delta\phi &= \arg p\\
a\cdot b &= |p|.
\end{align*}
    
The measured phase is $\Delta\phi\ \mathrm{mod}\ 2\pi$, but it can be unwrapped reliably if the phase front distortions are smooth. For the phase unwrapping we linearly extrapolate two neighboring data points and add $n\cdot 2\pi,\ n\in \mathbb{Z}$, to the next point such that the difference to the extrapolated value is minimal (Fig.\ \ref{fig:calibration}(d)). We oversample the phase in a small region to make sure there cannot be phase gradients $>\pi$ at the starting point of this unwrapping procedure.
    
Knowing the phase and amplitude profile of the beam in the image plane, we can make local modifications to the light phase by changing the grating phase $\Delta\phi(x, y)$, and to the light amplitude by changing the slit width $a(x, y)$, to match any target profile. This way we can remove aberrations and create arbitrary beam shapes.

\section{Beam shaping performance} \label{sec:beamshaping}%PPJZ

\begin{figure}
\begin{center}
\includegraphics[width=0.9\linewidth]{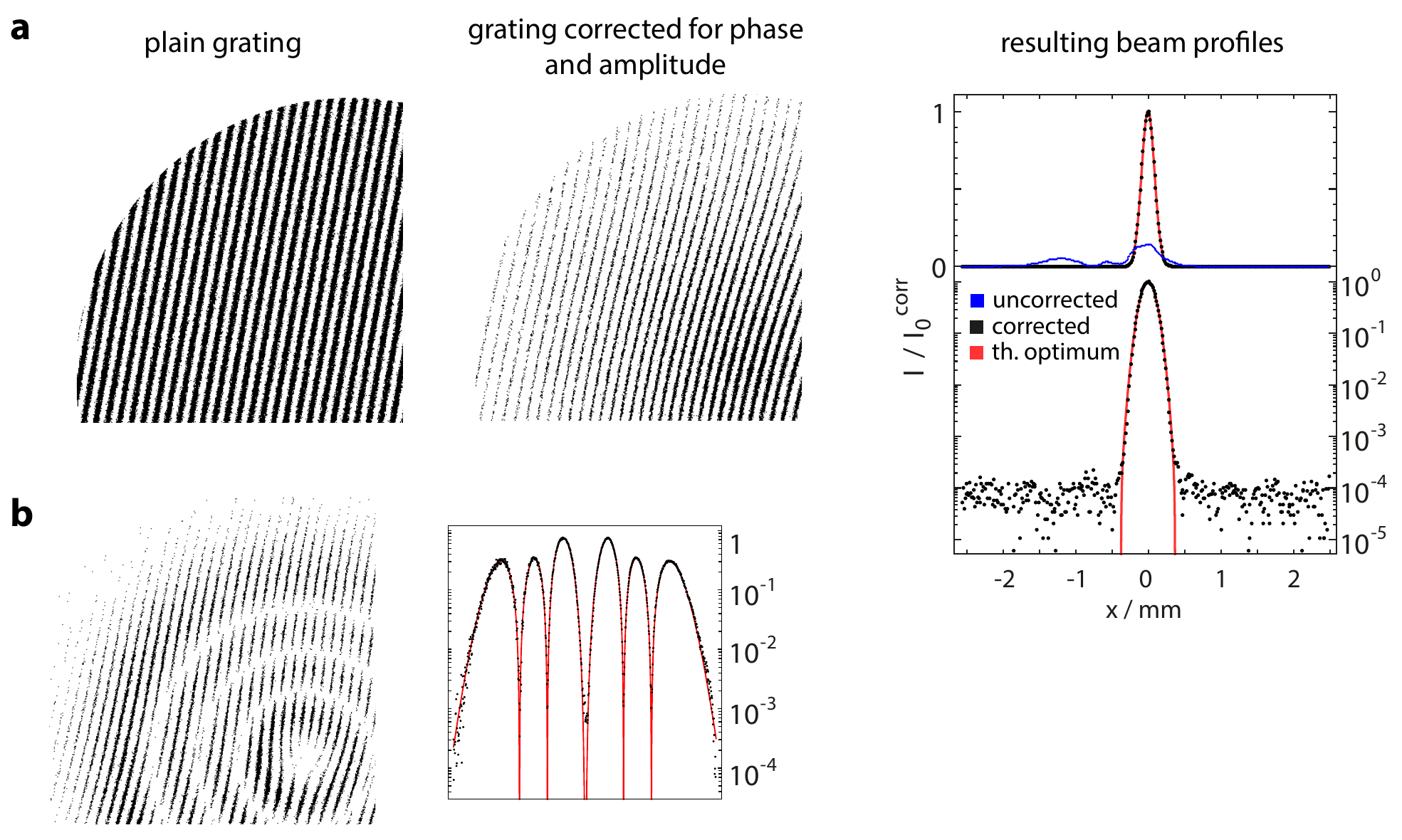}
\caption{High-precision beam shaping. (a) A diffraction grating is the basis of every DMD pattern. In an imaging system with strong aberrations, the unmodified pattern (left) leads to a corrugated beam shape (blue dots on the right). A phase- and amplitude-corrected pattern (center) cancels aberrations by bending the grating slits. This leads to the optimized profile (black dots), which very closely resembles the desired Gaussian profile over four orders of magnitude (red line). (b) By introducing phase vortices in addition to aberration and amplitude corrections, we generate clean higher-order modes, such as the Laguerre-Gauss $\mathrm{LG}_{33}$. We show a log--plot of the cross-section.}
\label{fig:beamshaping1}
\end{center}
\end{figure}   

In order to benchmark the holographic approach, we illuminate the DMD with a 532\,nm laser beam under an angle of $46^\circ$, for which the mirrors act as a blazed grating in the $+7^\mathrm{th}$ diffraction order at $-22^\circ$. Light in the first diffraction order of the DMD pattern is captured by a 1\,m~lens and focused onto a $10\,\mu$m pinhole, behind which a photodiode with low-noise amplifier measures the light intensity.

We run the calibration sequence to create a phase and amplitude map of the beam (Fig.\  \ref{fig:calibration}(e)). The phase data is interpolated to individual pixels and applied to eq.\ \ref{eqgrating} to compensate for aberrations. The amplitude map is used to calculate slit widths according to eq.\ \ref{eqslitwidth} to match the target profile. Even for highly corrugated initial beam profiles, application of the correction produces extremely precise beam shapes. Figure\ \ref{fig:beamshaping1}(a) shows a beam before and after correction. The corrected beam shows striking agreement with the Gaussian target profile and follows the theoretical prediction down to the noise floor at $10^{-4}$. Creating a phase map of the corrected beam shows residual phase errors below $\lambda/50$. The phase correction eliminates tilt and defocus just like the true optical aberrations, such that the beam is perfectly focused on the pinhole after the process. Conversely, it is also possible to add tilt and defocus purposefully to move the focus on the image plane or change the position of this plane. All beam profiles in this section are recorded by moving the beam on a grid over the pinhole and thus sampling its local intensity. We independently verified with a CCD camera at several exposure levels that this procedure correctly reproduces the beam profiles.

\begin{figure}
\begin{center}
\includegraphics[width=0.9\linewidth]{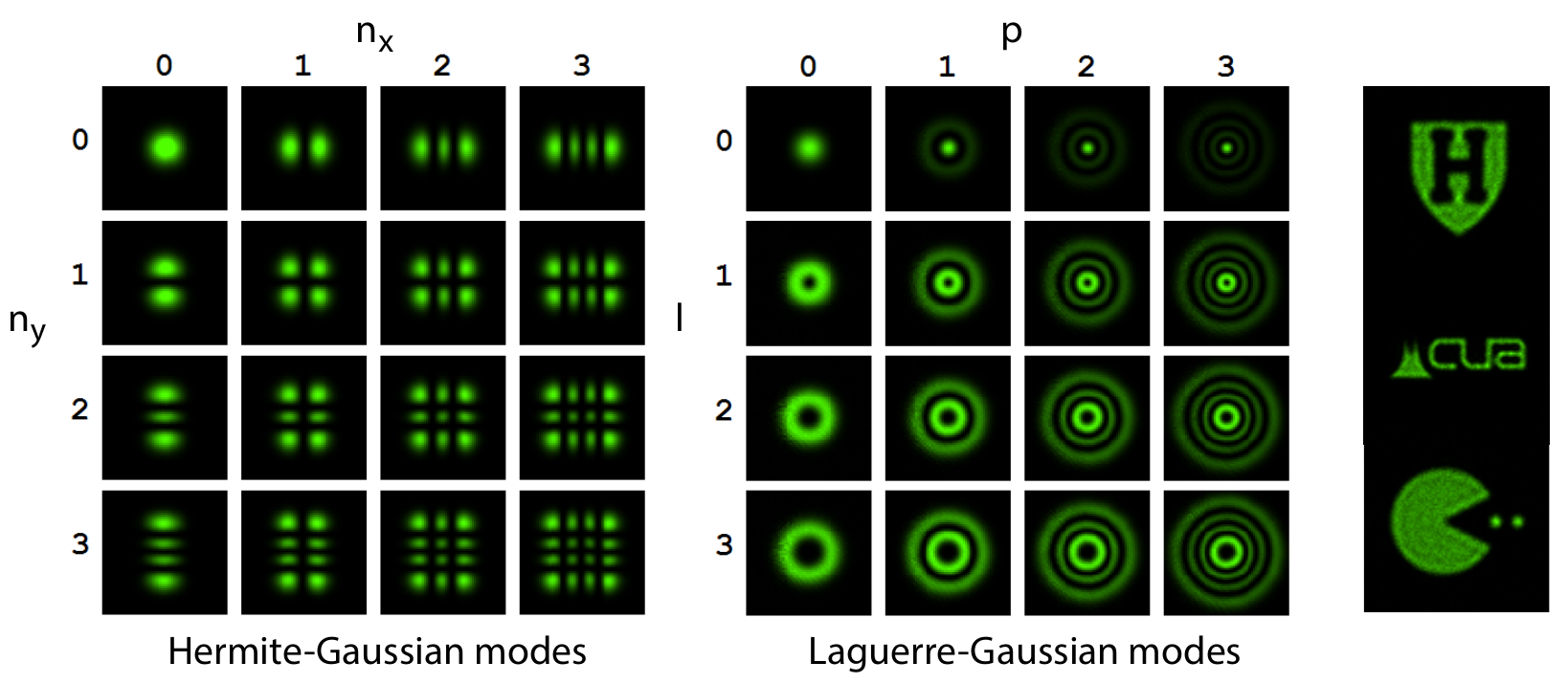}
\caption{Arbitrary beam shapes. Any shape can be projected by modulating its Fourier transform onto the DMD pattern. Combined with the aberration control, we create very clean higher-order Gaussian modes as well as more complicated, numerically defined structures.}
\label{fig:beamshaping2}
\end{center}
\end{figure}

Using the full control over amplitude and phase of the wavefront, extremely clean profiles can be created, including high-order Laguerre-Gaussian beams which require a phase vortex in the center (Fig.\ \ref{fig:beamshaping1}(b)). Analytic beam profiles and numerically defined shapes (Fig.\ \ref{fig:beamshaping2}) can be created in this fashion. The limiting factors are the finite resolution of the DMD and the binary nature of the amplitude modulation, which requires blocks of pixels for pseudo-grayscale pictures.
	
\section{Implementation in a quantum gas microscope}\label{sec:QGM_calibration} %PMP
We demonstrate the potential of holographic beam shaping techniques using a quantum gas microscope \cite{Preiss2015thesis}. Here, tailored beams can be used to address individual atoms in an optical lattice and to control the dynamics of many-body systems through the optical potential. The experimental setup has been described in previous work \cite{Bakr2009}. In brief, a two-dimensional quantum degenerate gas of \Rb resides in a square optical lattice with spacing $a=680\,$nm at the focus of a high-resolution imaging system with numerical aperture 0.8. The system realizes a Bose-Hubbard system with tunneling matrix element $J$ and on-site interaction $U$ \cite{Bakr2010}. We obtain single-atom and single-site resolved images of the many-body states using fluorescence imaging in deep optical lattices.

For this application, we use holographically generated beams to create arbitrary potentials on top of the optical lattice. The DMD is illuminated with coherent laser light at $\lambda = 760\,$nm, which is blue-detuned with respect to the $D1$ and $D2$ transition of \Rb and generates repulsive conservative optical potentials. The DMD is coupled into the microscope beam path using a dichroic, and is imaged onto the back focal plane of the objective, covering the full NA of 0.8. The effective aperture diameter on the DMD chip is 500 pixels (5.4\,mm).

We first calibrate all aberrations in the beam illuminating the DMD, the DMD itself, and the partial beam path using a photodiode as a point detector in an intermediate image plane, as outlined in section~\ref{sec:calibration}. In a second step, we calibrate aberrations in the remaining  part of the beam path to the atoms, including the high-NA objective. Because it is not possible to directly image the interference pattern or to place a photodiode in the plane of the atoms, we use the atoms themselves to probe aberrations in the microscope. Optimal alignment and focus onto the atoms in the lattice are thus guaranteed.

If used with resonant light and tightly trapped atoms, for example in ion traps, the DMD could be calibrated using the fluorescence rate of one (or several) atoms instead of a photodiode as a intensity point detector. Here, we instead use the response of the atoms to the conservative potential generated by far-detuned light to infer the phase differences between patches in the Fourier plane.

\begin{figure}
\begin{center}
\includegraphics[width=0.9\linewidth]{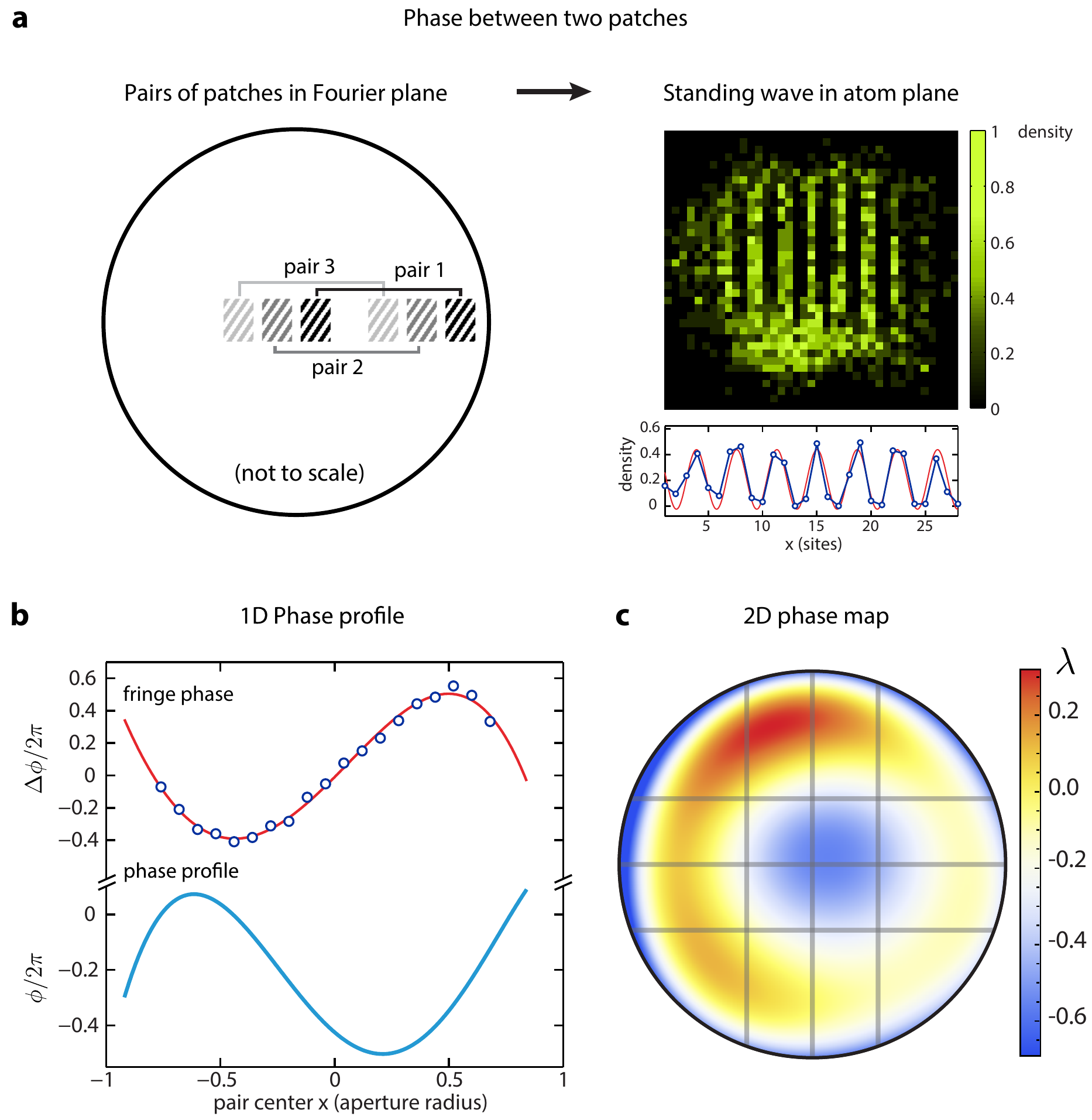}
\caption{Measuring aberrations in a quantum gas microscope. A Bose-Einstein condensate is used to map out interference patterns directly in the plane of the atoms. (a) Pairs of patches on the DMD (left) produce a standing wave in the atom plane, forcing the atoms to its minima. The measured atomic density distribution (right, one pixel corresponds to one lattice site) reveals the phase difference between the patches as the phase of the averaged density wave. (b) Translating the pair of patches across the Fourier plane, the changing standing wave phase reveals aberrations. Data points are measured fringe phases, red is a fit with the two shifted polynomials $\Delta \phi$ in eq.~(\ref{eq:deltaphi}). Light blue is the reconstructed phase profile $\phi$. The width of the plotted profile indicates the range of the aperture over which the reconstruction in valid. (c) A smooth phase map is interpolated from six measured phase profiles (gray lines). The dominant aberrations are spherical terms. Tilt and focus have been removed from the phase map.}
\label{fig:DMD_cal}
\end{center}
\end{figure}

We prepare a weakly interacting superfluid of $\unsim$400 atoms in a shallow two-dimensional lattice of depth 1\,$E_r$, where the recoil energy $E_r/\hbar= \angf 1240$\,Hz. We then adiabatically turn on the DMD beam while displaying a hologram of two patches spaced by 1/5 of the aperture (Fig.~\ref{fig:DMD_cal}). The atoms experience a repulsive standing-wave pattern of period 3.7 sites and arrange themselves in the minima of the DMD-generated potential. Images of the cloud directly reveal the phase difference between the two hologram patches on the DMD (Fig. \ref{fig:DMD_cal}). By averaging $\unsim10$ images and fitting the integrated density profile, we can determine the relative phase with a typical uncertainty of $\pm 0.2$\,rad. 
 
We map out the aberrations across the aperture by repeating the experiment with different pairs of patches. The spacing between the patches and hence the spacing of the interference pattern on the atoms remains fixed while the two patches are translated across the Fourier plane in steps of 1/25 of the aperture diameter.

Because the reference patch is translated with the sampling patch, the measured phase difference does not directly give the phase profile along the one-dimensional path. We model the profile with a $6^{\textrm{th}}$ order polynomial in the center of mass coordinate $x$ of the patch pair
\begin{equation}
\phi(x) = \sum_{i=0}^{6}a_i x^i
\end{equation}
The patches then sample the difference between two shifted polynomials
\begin{equation}
\Delta \phi(x)= \phi(x-l/2)-\phi(x+l/2)
\label{eq:deltaphi}
\end{equation}
where $l$ is the distance between the two patches. Fitting $\Delta \phi(x)$ to the measured phase difference between pairs along a cut through the Fourier plane yields the phase profile $\phi(x)$ up to an arbitrary offset $a_0$.

 We repeat this procedure along different paths and obtain six phase profile cuts through the Fourier plane (Fig. \ref{fig:DMD_cal}). The phase offset $a_0$ for each profile is adjusted to connect the cuts to a smooth two-dimensional surface. We interpolate the phase profile across the entire DMD aperture by fitting a two-dimensional surface parameterized by Zernike polynomials to the six phase profiles. We allow Zernike polynomials up to radial order 4 and azimuthal order 3. An example of a fitted phase map is shown in Fig. \ref{fig:DMD_cal}. Typical deviations from an expected flat wavefront after applying the correction are $\lambda / 6$ peak-to-peak, caused by  differences between the six measured phase profiles and the fitted two-dimensional surface. No additional amplitude map is obtained in the calibration to the atoms.

\section{Single-site addressing} \label{sec:QGM_addressing}%PMP
The calibrated beam-shaping setup in the quantum gas microscope can be used to generate arbitrary repulsive potentials with single-site resolution. The maximal deliverable power on the atoms is $900\,\mu\textrm{W}$, which for our setup corresponds to a Stark shift of ${V/\hbar=\angf 45 \,\textrm{MHz}}$ if focussed onto a single lattice site. Many potential configurations are possible, including box potentials, engineered disorder, or single-particle control of dynamics. Here we demonstrate single-atom addressing and few-body state preparation.

%The maximal deliverable power on the atoms is $900\,\mu\textrm{W}$, corresponding to an area-integrated Stark shift of ${V/\hbar=\angf 45 \,\textrm{MHz} \times (\textrm{site})^2}$ in our setup.
\begin{figure}
\begin{center}
\includegraphics[width=0.9\linewidth]{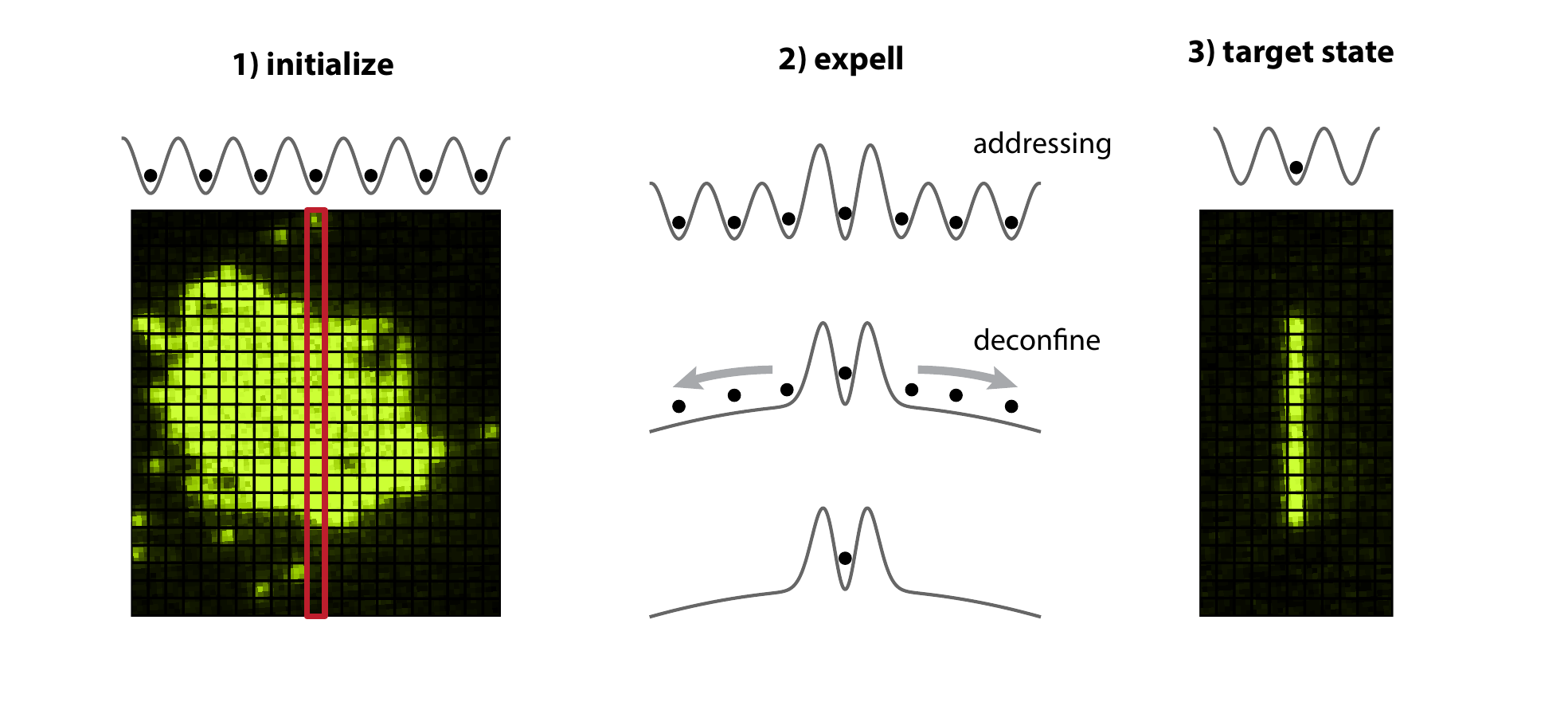}
\caption{Single-site addressing in an atomic Mott insulator: An addressing beam with a repulsive Hermite-Gauss profile is superimposed onto a Mott insulator in a deep optical lattice. A column of atoms is pinned in the minimum of the addressing beam while the transverse lattice is turned off and a deconfining Gaussian beam expels all other atoms. The pinned atoms are then loaded back into the optical lattice. By choosing higher orders of the addressing Hermite-Gauss beam, up to six adjacent columns can be prepared.}
\label{fig:DMD_cutting}
\end{center}
\end{figure}

\subsection{State initialization}
Few-particle Fock states in a lattice can be initialized by ``cutting" individual atoms from a low-entropy Mott insulator (Fig.~\ref{fig:DMD_cutting}). We prepare a Mott insulator with one particle per site in the atomic limit $U\gg J$ in a two-dimensional lattice with $V_x=V_y=45 E_r$. The DMD then superimposes a beam with Hermite-Gauss profile along in the transverse direction and a flattop profile along the longitudinal direction (length $10\,\mu$m). The distance between the peaks of the Hermite-Gauss profile is 930\,nm, with a typical peak depth of $25\,E_r$. We switch off the transverse lattice in the presence of a large ($40\,\mu$m waist) Gaussian, non-interfering deconfinement beam at 760\,nm. Only atoms in columns coinciding with the nodes of the Hermite-Gauss beam are retained, while all other atoms are expelled from the system within 40\,ms before the transverse lattice is ramped back on (Fig.~\ref{fig:DMD_cutting}). By choosing high-order Hermite-Gauss beams, we deterministically prepare up to six adjacent columns of atoms (length $\approx$10 sites). This cutting procedure retains atoms in the rows pinned by the DMD with a fidelity of $99(1)\%$, and the fidelity of preparing one atom per site is limited by the initialization fidelity of the Mott insulator of  typically $\geq 98\%$.

\subsection{Full number statistics}
\begin{figure}
\begin{center}
\includegraphics[width=0.9\linewidth]{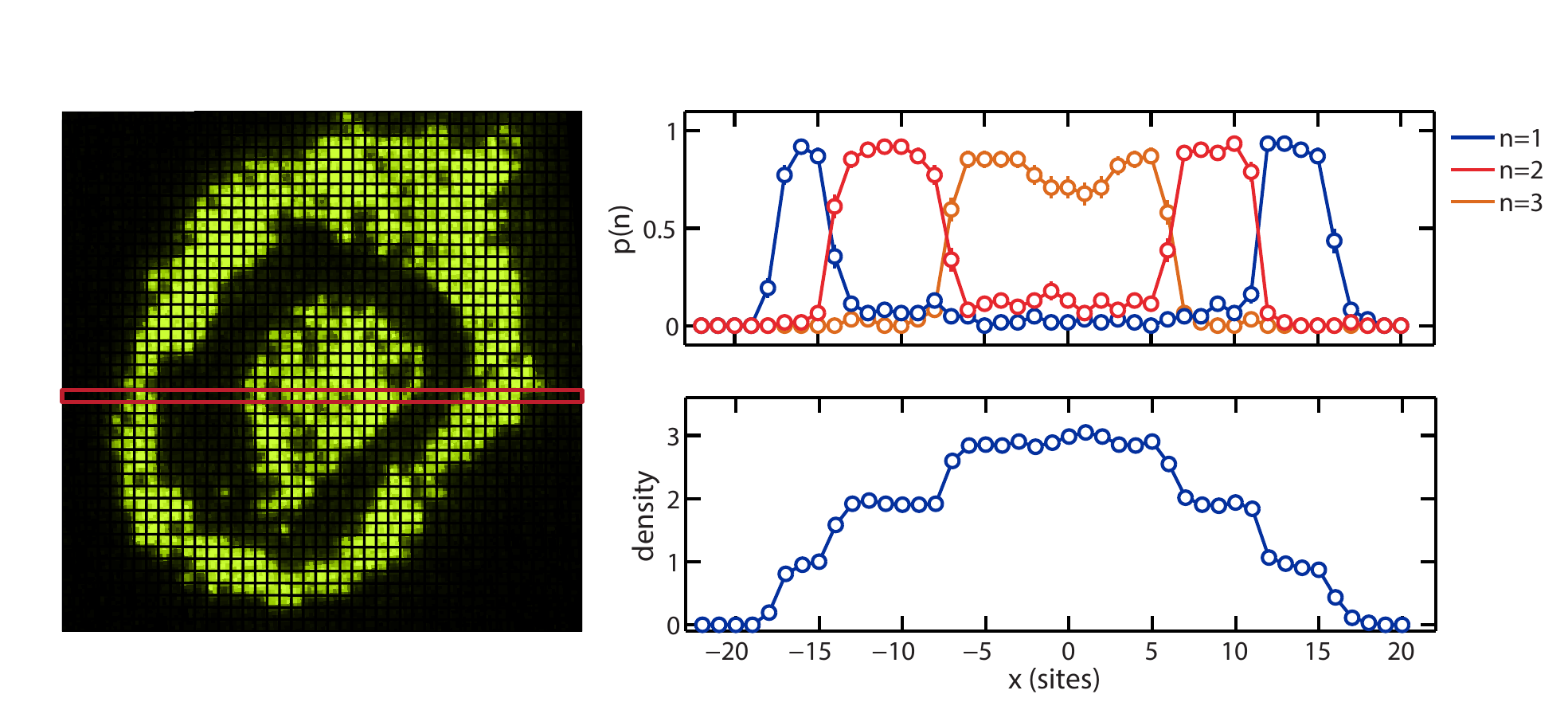}
\caption{Full counting statistics in one dimension. Left: A Mott insulator with up to $n=3$ atoms near the center appears as three concentric rings in parity-projecting fluorescence imaging. A single row or atoms (red box) can be cut from the Mott insulator and all atoms are detected after a short vertical expansion. Right: We obtain the full density profile and number statistics for each site in the one-dimensional profile.}
\label{fig:MI_n3_count}
\end{center}
\end{figure}

Our method of isolating well-defined subsystems from a many-body system can be used to circumvent parity projection inherent to \emph{in situ} fluorescence imaging \cite{Bakr2009, Sherson2010}. A large Mott insulator with up to three atoms per site appears as a ring of concentric bright and dark rings when imaged \emph{in situ}, corresponding to odd and even occupancy, respectively (Fig.~\ref{fig:MI_n3_count}). To avoid pairwise atom loss during imaging, we cut one row of atoms from the Mott insulator as above, emptying all other sites. We turn off the DMD addressing beam and release the atoms into one-dimensional tubes transverse to the cut, performing a $\unsim 5$\,ms free vertical expansion. The atoms delocalize over $\unsim100$ lattice sites during the time of flight, separating particles originating from the same site near the center of the Mott insulator. The atoms are then imaged individually without being lost to parity projection. Summing counts along transverse tubes yields the full number statistics of the one-dimensional cut through the two-dimensional system. Figure~\ref{fig:MI_n3_count} shows these results for an $n=3$ Mott insulator, directly revealing the wedding cake structure and site-resolved number distributions \cite{Preiss2015thesis}. The available field of view currently limits the transverse time of flight counting procedure to $n\leq4$ particles per site. Using higher laser powers in the addressing beam and a larger field of view, this technique could be applied to larger occupation numbers.

The full number statistics obtained through this procedure contain all the information to construct high-order number correlation functions between all sites within one cut. Such measurements are ideal for characterizing density correlations, for example in one-dimensional impurity problems \cite{Kantian2014}, or to detect entanglement through number fluctuations \cite{Smacchia2015}. In combination with a transverse magnetic field gradient \cite{Fukuhara2015}, it is possible to obtain fully spin-and number-resolved readout and study, for example, spin-charge separation \cite{Kleine2008}. 
 
\section{Conclusion}
We demonstrate a system for holographic beam shaping based on digital micromirror devices which achieves amplitude and phase control. Aberrations in the optical system can be measured and compensated to achieve precise alignment, focus, and diffraction-limited performance. Using the holographic beam shaping method with a quantum gas microscope, we demonstrate the creation of arbitrary potential landscapes and single-site addressing in an optical lattice. 

%Such techniques enable novel experiments, for example studies of few-particle dynamics \cite{Preiss2015} and the first direct measurements of entanglement entropy through the interference of two identical many-body states \cite{Islam2015}.

Our technique for aberration compensation is general, and can be of importance in a variety of experiments requiring precise optical wavefront control. Even better performance than demonstrated here can be achieved with DMD devices with more pixels, which are rapidly becoming available.

\section*{Acknowledgments}

We thank Sebastian Blatt for help in the early stages of the project and Adam Kaufman for discussions and a careful reading of the manuscript. Supported by grants from the Gordon and Betty Moore Foundations EPiQS Initiative (grant GBMF3795), NSF through the Center for Ultracold Atoms, the Army Research Office with funding from the DARPA OLE program and a MURI program, an Air Force Office of Scientific Research MURI program, and an NSF Graduate Research Fellowship (M.R.).

\end{document}